\documentclass{revtex4}

\usepackage[usenames,dvipsnames]{color}
\usepackage{amsfonts,amssymb,amsmath,amsthm}
\usepackage{bm}
\usepackage[a4paper]{geometry}
\usepackage{pdflscape}
\usepackage{graphicx}
\usepackage{mathrsfs}

\definecolor{darkred}{rgb}{.8,0,0}

\definecolor{darkblue}{rgb}{0,0,.7}

\newcommand{\eps}{\varepsilon}

\newcommand{\rev}[1]{\widetilde{#1}}
\newcommand{\blade}[2]{{#1}_1 \wedge \ldots \wedge {#1}_{#2}}

\newcommand{\uline}[1]{\underline{#1}}
\newcommand{\oline}[1]{\overline{#1}}

\theoremstyle{plain}
\newtheorem*{VarPrinc}{Variational principle}
\newtheorem*{CanEOM}{Canonical equations of motion}

\newtheorem*{TrPhysMot}{Transformation of physical motions}
\newtheorem*{SymHam}{Symmetry transformation}
\newtheorem*{Noether}{Noether theorem}

\begin{document}

\title{Hamiltonian constraint formulation of classical field theories}

\author{V\'{a}clav Zatloukal}

\email{zatlovac@fjfi.cvut.cz}

\homepage{http://www.zatlovac.eu}

\affiliation{\vspace{3mm}
Faculty of Nuclear Sciences and Physical Engineering, Czech Technical University in Prague, \\
B\v{r}ehov\'{a} 7, 115 19 Praha 1, Czech Republic \\
}

\affiliation{
Max Planck Institute for the History of Science, Boltzmannstrasse 22, 14195 Berlin, Germany
}

\begin{abstract}
Classical field theory is considered as a theory of unparametrized surfaces embedded in a configuration space, which accommodates, in a symmetric way, spacetime positions and field values. Dynamics is defined via the (Hamiltonian) constraint between multivector-valued generalized momenta, and points in the configuration space. Starting from a variational principle, we derive the local equations of motion, that is, differential equations that determine classical surfaces and momenta. A local Hamilton-Jacobi equation applicable in the field theory then follows readily. 
In addition, we discuss the relation between symmetries and conservation laws, and derive a Hamiltonian version of the Noether theorem, where the Noether currents are identified as the classical momentum contracted with the symmetry-generating vector fields. 
The general formalism is illustrated by two examples: the scalar field theory, and the string theory.

Throughout the article, we employ the mathematical formalism of geometric algebra and calculus, which allows us to perform completely coordinate-free manipulations.
\end{abstract}

\maketitle

%%%%%%%%%%%%%%%%%%%%%%%%%%%%%%%%%%%%%%
\section{Introduction}
%%%%%%%%%%%%%%%%%%%%%%%%%%%%%%%%%%%%%%

In non-relativistic mechanics, the trajectory of a particle is most commonly expressed as a function $x(t)$, which describes how the position of the particle evolves with time.
In relativistic mechanics, where space and time are treated in a symmetric way, the particle's trajectory is regarded as a sequence of spacetime events $(t,x)$. 

In field theory, the field configuration is usually regarded as a function $\phi(x)$, which describes how the values of the fields vary from point to point in the spacetime. However, the general relativity suggests \cite{RovelliQG} that the spacetime is a dynamical entity, and as such, it should be put with the fields on the same footing. Mathematically, instead of a function $\phi(x)$, one is therefore motivated to consider the respective graph, i.e., the collection of points $(x,\phi)$.

In this article, we develop the mathematical formalism for field theories proposed in \cite[Ch.~3]{RovelliQG} that treats time, space, and fields equally. All these entities are collectively called \emph{partial observables}, and they form a finite-dimensional \emph{configuration space}. Classical field theory studies correlations between the partial observables (called \emph{motions}), which have the form of surfaces embedded in the configuration space, and selects the \emph{physical (or classical) motions}, i.e., the ones that can be realized in nature.

Our dynamical description utilizes a multivector-valued momentum variable, which can be thought of as conjugated to the motion's tangent planes, thus generalizing the canonical momentum conjugated to the velocity vector in classical mechanics. Individual field theories are specified by a choice of the \emph{Hamiltonian} $H$, which is a function of the configuration space points $q$, and the momentum $P$. This Hamiltonian enters into a variational principle (Sec.~\ref{sec:VarPrinc}) via the so-called \emph{Hamiltonian constraint} $H(q,P) = 0$.

The aim of this article is to establish the Hamiltonian constraint formalism for the field theories as a viable, and even superior, alternative to the usual Lagrangian formalism. First, in Sec.~\ref{sec:CanEq}, we determine the canonical equations of motion, Eqs.~(\ref{CanEOM}), that follow from the variational principle. These equations generalize the Hamilton's canonical equations of motion of classical mechanics. In Sec.~\ref{sec:LocHJ}, we derive from Eqs.~(\ref{CanEOM}) a local Hamilton-Jacobi equation, Eq.~(\ref{HJeq}), which generalizes to the field theory the Hamilton-Jacobi equation of classical mechanics. (Our approach should be compared with Refs.~\cite{Kastrup,Rund}.) It is worth to emphasize that both, the canonical equations~(\ref{CanEOM}), and the Hamilton-Jacobi equation~(\ref{HJeq}), contain only partial, and not variational, derivatives. 

In Sec.~\ref{sec:Sym}, we study transformations of the configuration space, and specify the condition, Eq.~(\ref{SymCrit}), under which physical motions are mapped to physical motions. Such transformations are \emph{symmetries} of the physical system. The symmetries imply conservation laws through the Noether theorem \cite{Noether1918}, whose Hamiltonian version is derived in Sec.~\ref{sec:ConsLawsFromSym}. The corresponding conservation law (\ref{ConsLaw}) features a multivector-valued Noether current, obtained by contracting the momentum with a symmetry-generating vector field. 
%The latter observation provides a physical interpretation for the generalized momentum.

Two examples are provided to illustrate the universality of the presented formalism. The first example (Sec.~\ref{sec:ExScalar}) discusses the theory of a real multicomponent scalar field. It is shown that the canonical equations reproduce the De Donder-Weyl equations of motion \cite{DeDonder,Weyl,Kanat1999,Struckmeier}, and the local Hamilton-Jacobi equation reproduces the one invented by Weyl \cite{Weyl}, when the scalar field is regarded as a function defined on the spacetime. Moreover, we examine the symmetries, namely, spacetime translations and rotations, and rotations in the field space, and associate our multivector-valued Noether currents with the energy-momentum tensor and the angular-momentum tensor, and the standard vectorial Noether currents, respectively. 

In the second example (Sec.~\ref{sec:ExString}), we treat relativistic particles, strings, or higher-dimensional membranes, depending on the dimensionality of the motions. The configuration space is identified with the target space of the string theory, the motions are the worldsheets, and the Hamiltonian is essentially the simplest and most symmetric function of the momentum variable. The equations of motion have a simple geometric meaning, namely, they ensure that the mean curvature of the physical motion vanishes. In fact, this is exactly the condition that defines \emph{minimal surfaces} \cite{Osserman}. 
The Hamilton-Jacobi theory agrees with Ref. \cite{Nambu1980}. We also show that for nearly flat motions, the string theory yields the scalar field theory as a limiting case.

One more remark is in order before we start. All manipulations are performed in the mathematical language of geometric (or Clifford) algebra and calculus developed by D. Hestenes \cite{Hestenes} (see also Ref.~\cite{DoranLas}). We will assume that the reader is reasonably familiar with this language.
(A concise introduction into the geometric algebra techniques can be found in the appendices of Refs.~\cite{ZatlCanEOM} and \cite{ZatlSymConsLaws}. In these articles, we also provide a more detailed analysis of the subjects treated in the present article.)

\section{Variational principle} \label{sec:VarPrinc}
%%%%%%%%%%%%%%%%%%%%%%%%%%%%%%%%%%%%%%

Let us start with a set of partial observables that constitute a $D+N$-dimensional Euclidean configuration space $\mathcal{C}$. (An extension to pseudo-Euclidean spaces should be straightforward, but will not be discussed here.) A point $q$ in the configuration space, e.g., $q=(x,\phi)$, represents a simultaneous measurement of all partial observables.
To establish a \emph{physical} theory, one has to specify the correspondence between the partial observables and physical measuring devices, such as clocks, rulers, or instruments measuring the components of the field. In this article, we take such correspondence for granted, as we will only be concerned with the mathematical aspects of the theory.

Let us denote by $D$ the dimensionality of motions, i.e., submanifolds $\gamma$ of the configurations space $\mathcal{C}$. With $D=1$ one may study particle mechanics, with $D=2$ one can do the string theory or a field theory in two spacetime dimensions, and so on. We shall not consider systems with gauge invariance, for which the mathematical motion (the surface in $\mathcal{C}$) has higher dimensionality than the actual physical motion (the physical trajectory).

The tangent space of $\gamma$ at a point $q$ is spanned by $D$ linearly independent vectors $a_1,\ldots,a_D$, which are conveniently combined into a grade-$D$ multivector $\blade{a}{D}$. The normalized version of this multivector is called the \emph{unit pseudoscalar of $\gamma$}, and it is denoted by $I_\gamma$. In the terminology used in Ref. \cite[Ch.~6]{Frankel}, the function $I_\gamma(q)$ represents a $D$-dimensional distribution on $\mathcal{C}$, with $\gamma$ being its integral submanifold.

\begin{figure} 
\includegraphics[scale=1]{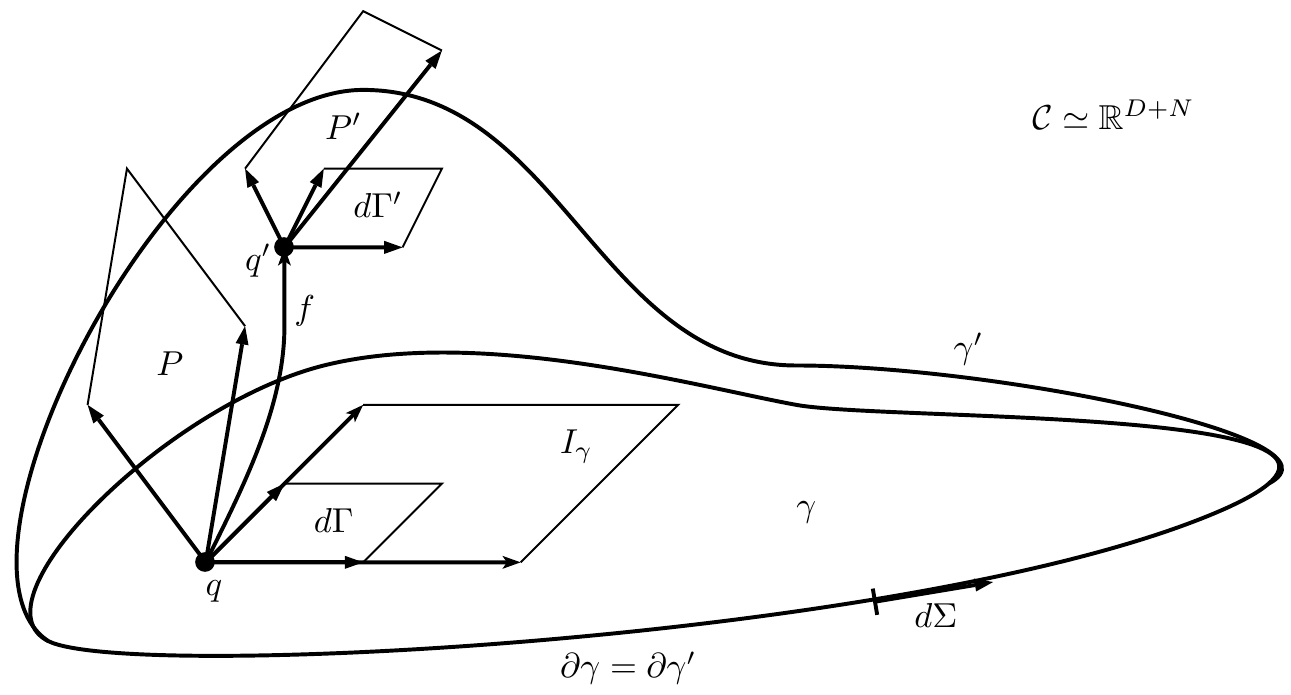}
\caption{Variational principle.}
\label{fig:VarPrinc}
\end{figure}

Fundamental for the following formulation of dynamics is the concept of the generalized momentum $P$, which is a grade-$D$ multivector defined at each point of $\gamma$ (see Fig.~\ref{fig:VarPrinc}). It serves as a quantity conjugated to $I_\gamma$, and in this sense it generalizes the canonical momentum of particle mechanics.

The Hamiltonian $H(q,P)$ is a generic function of positions and momenta, which is assumed to be scalar-valued. (A generalization to the case of multicomponent $H$ is straightforward.) 

The variational principle that determines the physical (or classical) motions of the field theory can now be stated as follows (cf. \cite[Ch.~3.3.2]{RovelliQG}): 
\begin{VarPrinc}
A surface $\gamma_{\rm cl}$ with boundary $\partial \gamma_{\rm cl}$ is a physical motion, if the couple $(\gamma_{\rm cl},P_{\rm cl})$ extremizes the (action) functional
\begin{equation} \label{Action}
\mathcal{A}[\gamma,P] = \int_\gamma P(q) \cdot d\Gamma(q)
\end{equation}
in the class of pairs $(\gamma,P)$, for which $\partial\gamma = \partial\gamma_{\rm cl}$, and for which $P$, defined along $\gamma$, obeys the Hamiltonian constraint
\begin{equation} \label{HamConstraint}
H(q,P(q)) = 0 ~~~~~ \forall q \in \gamma .
\end{equation}
\end{VarPrinc}

The integral in (\ref{Action}) is defined in \cite[Ch.~7]{Hestenes} (see also Ref.~\cite{Sobczyk}) without resorting to any parametrization of the surface $\gamma$. 
The inner product between the momentum $P$ and the oriented surface element $d\Gamma$ replaces the differential form $\theta = p_{j_1 \ldots j_D} dq^{j_1} \wedge \ldots \wedge dq^{j_D}$ used in Ref.~\cite[Ch.~3.3.2]{RovelliQG}. There, the integral is taken over a submanifold of the bundle of $D$-forms over $\mathcal{C}$. Since we hesitate to work in spaces that mix points $q$ and multivectors $P$, we prefer to integrate over surfaces in $\mathcal{C}$, and treat the momentum as a field defined along these surfaces.

%%%%%%%%%%%%%%%%%%%%%%%%%%%%%%%%%%%%%%
\section{Canonical equations of motion} \label{sec:CanEq} 
%%%%%%%%%%%%%%%%%%%%%%%%%%%%%%%%%%%%%%

We will now derive the equations of motion that follow from the variational principle. For this purpose, we incorporate the Hamiltonian constraint (\ref{HamConstraint}) into the action (\ref{Action}) by means of a scalar Lagrange multiplier $\lambda$. The augmented action is a functional
\begin{equation} \label{ActionAugm}
\mathcal{A}[\gamma,P,\lambda] 
= \int_\gamma \left[ P(q) \cdot d\Gamma(q) - \lambda(q) H(q,P(q)) \right] ,
\end{equation}
where $\lambda$ is, in fact, an infinitesimal quantity comparable with $|d\Gamma|$, the magnitude of $d\Gamma$.

The varied action $\mathcal{A}[\gamma',P',\lambda']$ is an integral over a new surface $\gamma'$, with new functions $P'$ and $\lambda'$ defined along $\gamma'$ (see Fig.~\ref{fig:VarPrinc}). Let 
\begin{equation}
f(q) = q + \delta q(q) 
\end{equation}
be the infinitesimal diffeomorphism mapping the surface $\gamma$ to $\gamma'$, i.e., $\gamma'=\{q'=f(q)\,|\,q \in \gamma \}$, and let us denote by 
\begin{equation}
\delta P(q) \equiv P'(f(q)) - P(q) 
~~~~~{\rm and}~~~~~ 
\delta \lambda(q) \equiv \lambda'(f(q)) - \lambda(q)
\end{equation}
the variations of the momentum and the Lagrange multiplier, respectively.

The infinitesimal variation of the action (\ref{ActionAugm}), $\delta\mathcal{A} \equiv \mathcal{A}[\gamma',P',\lambda'] - \mathcal{A}[\gamma,P,\lambda]$, is then given by
\begin{equation}
\delta\mathcal{A} 
= \int_\gamma \left[ P'(f(q)) \cdot \underline{f}(d\Gamma(q);q) \!-\! \lambda'(f(q)) H \big(f(q),P'(f(q)) \big) \right]
- \int_\gamma \left[ P(q) \cdot d\Gamma(q) \!-\! \lambda(q) H(q,P(q)) \right] ,
\end{equation}
where we have employed the integral substitution theorem (see \cite[Ch.~7-5]{Hestenes}) to transform the integral over $\gamma'$ to an integral over $\gamma$. For the infinitesimal diffeomorphism $f$, the outermorphism mapping $\underline{f}$ that specifies the transformation rule for multivectors, is given by Formula~(\ref{GCinfsmTr}). Therefore, to the first order in $\delta q$, $\delta P$, and $\delta \lambda$, we find
\begin{align} \label{ActionVar}
\delta\mathcal{A}
&= \int_\gamma \left[ 
(P + \delta P) \cdot \big(d\Gamma + (d\Gamma \cdot \partial_q) \wedge \delta q \big) - (\lambda + \delta\lambda) H(q + \delta q, P + \delta P)
- P \cdot d\Gamma + \lambda\, H(q,P) 
\right]
\nonumber\\
&\approx 
\int_\gamma \left[
- \delta\lambda \, H(q,P) 
+ \delta P \cdot \big(d\Gamma - \lambda \, \partial_P H(q,P) \big)
- \lambda \, \delta q \cdot \dot{\partial}_q H(\dot{q},P)
+ P \cdot \big((d\Gamma \cdot \partial_q) \wedge \delta q \big)
\right] ,
\end{align}
where $\partial_q$ is the \emph{vector derivative} with respect to a point in $\mathcal{C}$ \cite[Ch.~2-1]{Hestenes}, and $\partial_P$ is the \emph{multivector derivative} with respect to the momentum multivector $P$  \cite[Ch.~2-2]{Hestenes}.
The ``overdot" notation is used to indicate the scope of the differential operator $\partial_q$. Without an overdot, any differential operator is supposed to act on the functions that stand to its right.

The last term in Eq.~(\ref{ActionVar}) can be rewritten with a help of the \emph{Fundamental theorem of geometric calculus} \cite[Ch.~7-3]{Hestenes},
\begin{equation} \label{PerPartes}
\int_\gamma P \cdot \big((d\Gamma \cdot \partial_q) \wedge \delta q \big)
= \int_{\partial\gamma} P \cdot (d\Sigma \wedge \delta q)
- \int_\gamma \dot{P} \cdot \big((d\Gamma \cdot \dot{\partial}_q) \wedge \delta q \big) ,
\end{equation}
where $d\Sigma$ is the oriented surface element on the boundary $\partial\gamma$. Now, the first term on the right-hand side of this equation vanishes, since we assume that $\gamma$ and $\gamma'$ have a common boundary, i.e., $\delta q|_{\partial\gamma}=0$. As concerns the second term, for $D=1$, $d\Gamma \cdot \partial_q$ is algebraically a scalar, and so the integrand is readily reshuffled,
\begin{equation}
\dot{P} \cdot \big((d\Gamma \cdot \dot{\partial}_q) \wedge \delta q \big)
= \delta q \cdot \, (d\Gamma \cdot \partial_q P) .
\end{equation}
(Mind the priority of the inner product ``$\cdot$", and the outer product ``$\wedge$" before the geometric product, which is denoted by an empty symbol.)
For $D>1$, we may employ some basic geometric algebra identities to find
\begin{equation}
\dot{P} \cdot \big((d\Gamma \cdot \dot{\partial}_q) \wedge \delta q \big)
= \big( \dot{P} \cdot (d\Gamma \cdot \dot{\partial}_q) \big) \cdot \delta q
= (-1)^{D-1} \delta q \cdot \big( (d\Gamma \cdot \partial_q) \cdot P \big) .
\end{equation}
The two cases have to be treated separately due to the definition of the inner product adopted in Ref.~\cite[Ch.~1]{Hestenes}.

After these rearrangements, we arrive at our final expression for the variation of the action,
\begin{equation}
\delta\mathcal{A}
\approx 
\int_\gamma \left[
- \delta\lambda \, H(q,P) 
+ \delta P \cdot \big( d\Gamma - \lambda \, \partial_P H(q,P) \big)
+ \delta q \cdot \left(
(-1)^D (d\Gamma \cdot \partial_q) \cdot P
- \lambda \, \dot{\partial}_q H(\dot{q},P)
 \right)
\right] ,
\end{equation}
which holds for $D>1$, while the case $D=1$ is obtained simply by replacing $(d\Gamma \cdot \partial_q) \cdot P$ with $d\Gamma \cdot \partial_q P$.
The requirement that $\delta\mathcal{A}$ vanish for all $\delta P$, $\delta q$, and $\delta\lambda$ yields the following
\begin{CanEOM}
Physical motions $\gamma_{\rm cl}$ are obtained by solving the system of differential equations
\begin{subequations} \label{CanEOM}
\begin{align} 
\label{CanEOM1}
\lambda \, \partial_P H(q,P) &= d\Gamma , 
\\ \label{CanEOM2}
(-1)^D \lambda \, \dot{\partial}_q H(\dot{q},P) 
&= \begin{cases}
d\Gamma \cdot \partial_q P & ~~{\rm for}~ D=1 \\
(d\Gamma \cdot \partial_q) \cdot P &  ~~{\rm for}~ D>1 ,
\end{cases}
\\
\label{CanEOM3} 
H(q,P) &= 0 .
\end{align}
\end{subequations}
\end{CanEOM}
(We use the adjective ``canonical", because these equations generalize the Hamilton's canonical equations of motion of classical mechanics \cite{ZatlCanEOM}.)

The first canonical equation (\ref{CanEOM1}) furnishes a relation between the momentum $P$, and the tangent planes of $\gamma$, represented by the oriented surface element $d\Gamma$. It asserts that the multivector derivative $\partial_P H$, which is a grade-$D$ multivector, is proportional to $d\Gamma$, with the proportionality constant equal to $\lambda$. Note that one can always divide $\lambda$ and $d\Gamma$ by the magnitude $|d\Gamma|$ to free Eqs.~(\ref{CanEOM}) from infinitesimal quantities.

The second canonical equation (\ref{CanEOM2}) describes how the momentum multivector $P$ changes as it slides along the surface $\gamma$. It is important to note that $P$ is being differentiated, effectively, only in the directions parallel to $\gamma$, as a consequence of the inner product between the surface element $d\Gamma$, and the vector derivative $\partial_q$. Moreover, the ``overdot" on the left-hand side  assures that only the explicit dependence of $H$ on $q$ is being differentiated, not the dependence through $P(q)$.

The last canonical equation (\ref{CanEOM3}) is simply the Hamiltonian constraint (\ref{HamConstraint}). Let us remark that had we started with several constraints $H_j(q,P)=0$ in the variational principle, we would have introduced the corresponding number of Lagrange multipliers $\lambda_j$, and, consequently, the canonical equations would contain the terms $\sum_j\lambda_jH_j$ instead of $\lambda\,H$.

\section{Local Hamilton-Jacobi theory} \label{sec:LocHJ}
%%%%%%%%%%%%%%%%%%%%%%%%%%%%%%%%%%%%%%

One method to deal with the canonical equations is the following. Suppose $P(q)$ obeys the Hamiltonian constraint
\begin{equation}
H(q,P(q)) = 0 
\end{equation}
in some $D+N$-dimensional region in the configuration space $\mathcal{C}$. By differentiation, we obtain
\begin{equation}
\dot{\partial}_q H(\dot{q},P(q))
+ \dot{\partial}_q \dot{P}(q) \cdot \partial_P H(q,P(q)) = 0 ,
\end{equation}
and using the first canonical equation (\ref{CanEOM1}), we find that
\begin{equation}
\lambda \, \dot{\partial}_q H(\dot{q},P(q))
= - \dot{\partial}_q \dot{P}(q) \cdot d\Gamma .
\end{equation}
The right-hand side may be recast, using the identities (1.42) and (1.43) from Ref.~\cite{Hestenes}, in the form
\begin{equation} \label{HJeom2}
\lambda \, \dot{\partial}_q H(\dot{q},P(q)) =
\begin{cases}
d\Gamma \cdot \big( \partial_q \wedge P(q) \big)
- d\Gamma \cdot \partial_q P(q) & ~~{\rm for}~ D=1
\\
(-1)^{D-1} d\Gamma \cdot \big( \partial_q \wedge P(q) \big)
+ (-1)^D (d\Gamma \cdot \partial_q) \cdot P(q) & ~~{\rm for}~ D>1 .
\end{cases}
\end{equation}

Now, we observe that if
\begin{equation}
\partial_q \wedge P(q) = 0 ,
\end{equation}
then Eq.~(\ref{HJeom2}) coincides with the second canonical equation (\ref{CanEOM2}), which is then automatically fulfilled. The momentum field that satisfies this condition can be expressed, at least locally, as $P(q) = \partial_q \wedge S(q)$, where $S$ is a multivector of grade $D-1$ (cf. the relation between closed and exact differential forms). The canonical equations (\ref{CanEOM}) are then reduced to two equations:
\begin{equation} \label{HJeqGamma}
\lambda \, \partial_P H(q,\partial_q \wedge S) = d\Gamma ,
\end{equation}
and the \emph{local Hamilton-Jacobi equation}
\begin{equation} \label{HJeq}
H(q,\partial_q \wedge S) = 0 .
\end{equation}

If we succeed in finding a solution of Eq. (\ref{HJeq}), we can plug it into Eq. (\ref{HJeqGamma}), which then defines a distribution of the tangent planes of a classical motion surface $\gamma_{\rm cl}$. This distribution can be integrated to yield the surface itself, provided certain integrability conditions are met (see \cite[Ch.~6.1]{Frankel}).

If we find a whole family of solution $S(q;\alpha)$, parametrized by a continuous parameter $\alpha$, then, by differentiating Eq. (\ref{HJeq}) with respect to $\alpha$, and substituting Eq. (\ref{HJeqGamma}), we obtain the relation
\begin{equation} \label{HJparamSol}
0 = \lambda \, \partial_\alpha H(q,\partial_q \wedge S) 
= \lambda \, \dot{\partial}_\alpha (\partial_q \wedge \dot{S}) \cdot \partial_P H(q,\partial_q \wedge S)
= d\Gamma \cdot \big(\partial_q \wedge (\partial_\alpha S) \big) .
\end{equation}
Now, for $D=1$, the Hamilton-Jacobi function $S$ is scalar-valued, and we obtain
\begin{equation} \label{HJconserved}
d\Gamma \cdot \partial_q (\partial_\alpha S) = 0 
~~~\Rightarrow~~~
\partial_\alpha S(q;\alpha) = \beta ~~~~~\forall q \in \gamma_{\rm cl} ,
\end{equation}
for some constant $\beta$, meaning that the quantity $\partial_\alpha S(q;\alpha)$ is conserved along a physical motion. Finding $N$ such parameters $\alpha$ (recall that the dimension of the configuration space is now $1+N$), the physical motion $\gamma_{\rm cl}$ can be determined from the set of constraints between the partial observables,
\begin{align}
\partial_{\alpha_1} S(q;\alpha_1,&\ldots,\alpha_N) = \beta_1 \nonumber \\
&\vdots \nonumber \\
\partial_{\alpha_N} S(q;\alpha_1,&\ldots,\alpha_N) = \beta_N .
\end{align}
Of course, we assume that the $N$ constraints are independent, i.e., that the gradients $\partial_q (\partial_{\alpha_1} S), \ldots, \partial_q (\partial_{\alpha_N} S)$ are, at every point, linearly independent vectors. 
%In Example \ref{sec:ExString} we will illustrate the Hamilton-Jacobi method with the case of a relativistic particle. 

When $D>1$, Eq.~(\ref{HJparamSol}) can be rearranged, and integrated using the fundamental theorem of geometric calculus, \begin{equation} \label{HJcontEq}
(d\Gamma \cdot \partial_q) \cdot (\partial_\alpha S) = 0 
~~~\Rightarrow~~~
\int_{\bar{\gamma}_{\rm cl}} (d\Gamma \cdot \partial_q) \cdot (\partial_\alpha S)
= \int_{\partial \bar{\gamma}_{\rm cl}} d\Sigma \cdot (\partial_\alpha S)
= 0 ,
\end{equation}
where $\bar{\gamma}_{\rm cl}$ is an arbitrary $D$-dimensional subset of $\gamma_{\rm cl}$ (a ``patch" on $\gamma_{\rm cl}$). Eqs.~(\ref{HJconserved}) and (\ref{HJcontEq}) express conservation laws for the conserved quantities $\partial_\alpha S$ (see the Noether theorem, Eq.~(\ref{ConsLaw}), below). 

A remark is in order before we close this section. In classical particle mechanics, one of the solutions of the Hamilton-Jacobi equation is the action along a classical trajectory, regarded as a function of one of the endpoints. In the field theory, the classical action may be viewed as a functional of the boundary $\partial\gamma_{\rm cl}$. Some authors (e.g., \cite[Ch.~3.3.4]{RovelliQG}) have therefore considered a variational differential equation that describes how the classical action changes under variations of the boundary, using also the name ``Hamilton-Jacobi equation". Note that Eq. (\ref{HJeq}) is substantially different from this kind of approaches, for it contains only partial, not variational, derivatives. This is why we call it ``local Hamilton-Jacobi equation". A local Hamilton-Jacobi theory is also treated, e.g., in Refs. \cite{Kastrup} and \cite{Rund}.

%%%%%%%%%%%%%%%%%%%%%%%%%%%%%%%%%%%%%%
\section{Symmetries in the Hamiltonian approach}
%%%%%%%%%%%%%%%%%%%%%%%%%%%%%%%%%%%%%%
\label{sec:Sym}

In this section, we will study transformations of the configuration space $\mathcal{C}$ of partial observables, and identify among them the symmetries of a physical system. 

\begin{figure} 
\includegraphics[scale=1]{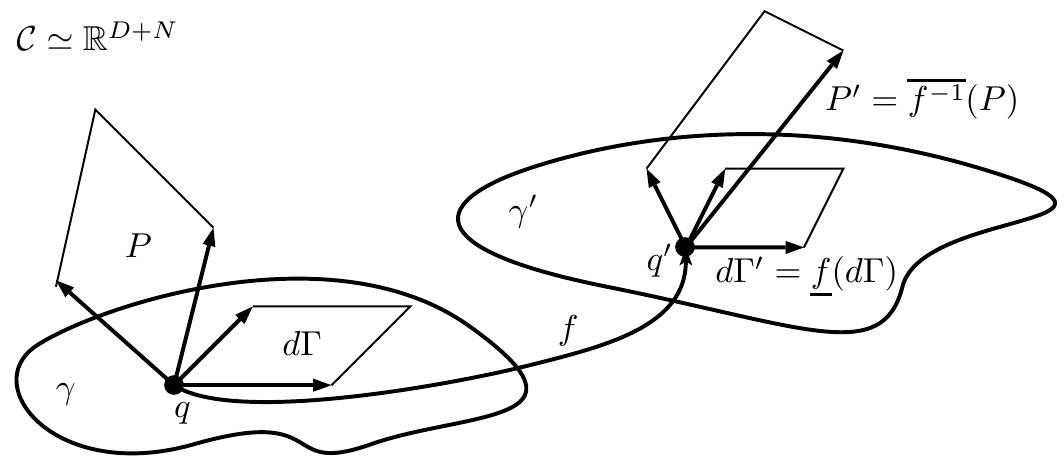}
\caption{The transformation of motions, surface elements, and the momenta under a diffeomorphism $f$.}
\label{fig:Transf}
\end{figure}

A transformation of $\mathcal{C}$ is expressed mathematically as a diffeomorphism $f:\mathcal{C}\rightarrow\mathcal{C}$ (see Fig.~\ref{fig:Transf}). 
It maps a surface $\gamma$ to another surface 
\begin{equation}
\gamma'=\{q'=f(q)\,|\,q \in \gamma \} ,
\end{equation}
whose boundary $\partial\gamma'$ may differ from $\partial\gamma$.
The surface elements on $\gamma$ and $\gamma'$ are related by the induced outermorphism $\underline{f}$, 
\begin{equation}
d\Gamma'(q')~=~\underline{f}(d\Gamma(q);q) .
\end{equation}
(Transformations and induced mappings within the framework of geometric calculus are introduced in Appendix~\ref{sec:GAGC}, and thoroughly discussed in \cite[Ch.4-5]{Hestenes}.) Note that $f$ is an \emph{active} transformation, a mapping between the points of the configuration space $\mathcal{C}$. In a dual picture, one could consider the \emph{passive} transformations, i.e., changes of the coordinates on $\mathcal{C}$. Since we are working completely without coordinates, all transformations are viewed as active.

The relation between the momentum fields on $\gamma$ and $\gamma'$ is established by demanding that the inner product $P \cdot d\Gamma$, and hence the action (\ref{Action}), be invariant under $f$. This is achieved by postulating the transformation rule
\begin{equation} \label{TrP}
P' = \overline{f^{-1}}(P;q) .
\end{equation}
The invariance of the action then implies the following:
\begin{TrPhysMot}
Consider an arbitrary diffeomorphism $f:\mathcal{C}\rightarrow\mathcal{C}$. If $\gamma_{\rm cl}$ is a physical motion of a system with Hamiltonian $H$, then
\begin{equation} \label{TrPhysMotion}
\gamma'_{\rm cl} = \{q'=f(q)\,|\,q \in \gamma_{\rm cl} \}
\end{equation}
is a physical motion of a system with Hamiltonian $H'$, defined by
\begin{equation} \label{TrHam}
H'(q',P') = H(q,P) ,
\end{equation}
where $P'=\overline{f^{-1}}(P;q)$.
\end{TrPhysMot}
(An explicit proof of this claim on the level of canonical equations of motion is provided in Ref.~\cite{ZatlSymConsLaws}.)

%Note that the definition of $P'$, Eq.~(\ref{TrP}), introduces an additional $q$-dependency into the Hamiltonian $H'$ as compared to $H$, due to the $q$-dependency of $\overline{f^{-1}}$.

We call the transformation $f$ a \emph{symmetry}, if it maps physical motions to physical motions of the same physical system. This is the case when $H$ and $H'$ coincide, i.e., when
\begin{equation}
H'(q',P') = H(q',P') .
\end{equation}
As an immediate consequence of definition (\ref{TrHam}), we therefore obtain:
\begin{SymHam}
A transformation $f$ is a symmetry of a physical system described by the Hamiltonian $H$ (or, in short, a symmetry of $H$), if
\begin{equation} \label{SymCrit}
H(f(q),\overline{f^{-1}}(P;q)) = H(q,P) .
\end{equation}
For infinitesimal transformations $f(q) = q + \eps\, v(q)$, $\eps \ll 1$, determined by a vector field $v$, Eq.~(\ref{SymCrit}) takes the form
\begin{equation} \label{SymCritInfsm}
v \cdot \dot{\partial}_q H(\dot{q},P)
- \big( \dot{\partial}_q \wedge (\dot{v} \cdot P) \big) \cdot \partial_P H(q,P)
= 0 .
\end{equation}
\end{SymHam}

Eq.~(\ref{SymCritInfsm}) is obtained from Eq.~(\ref{SymCrit}) by a straightforward application of the infinitesimal version of the transformation rule (\ref{TrP}), Eq.~(\ref{GCinfsmInvTr}).

More rigorously, the infinitesimal transformation arises from a one-parameter group of transformations $f_\tau(q)$ in the small-$\tau$ limit, when we can approximate
\begin{equation}
f_\tau(q) \approx q + \tau v(q)
~~~,~~~ v(q) = \partial_\tau f_\tau(q)|_{\tau=0} .
\end{equation}
Conversely, to any vector field $v(q)$ corresponds a flow $f_\tau(q)$, which can be regarded as a group of transformations parametrized by $\tau$. An explicit formula is provided by the \emph{Lie series} \cite[Ch.~1.3]{Olver},
\begin{equation} \label{LieSeries}
f_\tau(q) = e^{\tau v \cdot \partial_q} q 
= q + \tau v + \frac{\tau^2}{2!} (v \cdot \partial_q) v + \ldots .
\end{equation}

%%%%%%%%%%%%%%%%%%%%%%%%%%%%%%%%%%%%%%
\section{Conservation laws from symmetries}
%%%%%%%%%%%%%%%%%%%%%%%%%%%%%%%%%%%%%%
\label{sec:ConsLawsFromSym}

The symmetries of a physical system are imprinted in its Hamiltonian function $H(q,P)$, and can be explored by analysing Eqs.~(\ref{SymCrit}) or (\ref{SymCritInfsm}) without any reference to the equations of motion. 

However, when the system is assumed to follow a classical trajectory, then the symmetries induce \emph{conservation laws}. This fact is derived almost instantly in the Hamiltonian constraint formalism.
Substituting canonical equations (\ref{CanEOM1}) and (\ref{CanEOM2}), respectively, into the first and the second term in Eq.~(\ref{SymCritInfsm}), we find (for $D>1$)
\begin{equation}
(-1)^D v \cdot \big( (d\Gamma \cdot \partial_q) \cdot P \big)
- \big( \dot{\partial}_q \wedge (\dot{v} \cdot P) \big) \cdot d\Gamma
 = 0 ,
\end{equation}
which can be readily rearranged,
\begin{equation}
(d\Gamma \cdot \dot{\partial}_q) \cdot (\dot{P} \cdot v) +
(d\Gamma \cdot \dot{\partial}_q) \cdot (P \cdot \dot{v})
= 0 ,
\end{equation}
and finally combined into one term to yield the equation
\begin{equation}
(d\Gamma \cdot \partial_q) \cdot (P \cdot v)
= 0 .
\end{equation}
The derivation for the case $D=1$ is fully analogous. 

Let us summarize the above considerations in the following Hamiltonian version of the celebrated
\begin{Noether}
If $f(q)=q+\eps v(q)$ is an infinitesimal symmetry of $H$, i.e., if Eq.~(\ref{SymCritInfsm}) holds, then the solutions of the canonical equations of motion (\ref{CanEOM}) satisfy the conservation law 
\begin{align} \label{ConsLaw}
d\Gamma \cdot \partial_q \, (P \cdot v)
&= 0  \hspace{10mm}{\rm for}~ D=1  \nonumber\\
(d\Gamma \cdot \partial_q) \cdot (P \cdot v)
&= 0 \hspace{10mm}{\rm for}~ D>1 .
\end{align}
\end{Noether}

The quantities that obey conservation laws play distinguished role in physics. The Noether theorem therefore grants a special status to the $D-1$-vector $P \cdot v$, and clearly displays the importance of the momentum multivector $P$ not only in particle mechanics, but also in the classical field theory.

The integral form of the conservation laws is obtained, analogously to Sec.~\ref{sec:LocHJ}, by integrating Eq.~(\ref{ConsLaw}) over an arbitrary connected $D$-dimensional subset $\bar{\gamma}_{\rm cl}$ of a physical motion $\gamma_{\rm cl}$, and by employing the fundamental theorem of geometric calculus. For $D=1$, we obtain
\begin{equation} \label{ConsLawIntD1}
P(q_2) \cdot v(q_2) - P(q_1) \cdot v(q_1) 
= 0 ,
\end{equation}
where $q_1$, $q_2$ are the endpoints of the curve $\bar{\gamma}_{\rm cl}$, whereas for $D>1$, we find
\begin{equation} \label{ConsLawIntDD}
\int_{\partial \bar{\gamma}_{\rm cl}} d\Sigma \cdot (P \cdot v)
= 0 ,
\end{equation}
where $d\Sigma$ is the oriented infinitesimal surface element of the boundary $\partial \bar{\gamma}_{\rm cl}$.

%
%%%%%%%%%%%%%%%%%%%%%%%%%%%%%%%%%%%%%%%
%\section{Examples} \label{sec:Examples}
%%%%%%%%%%%%%%%%%%%%%%%%%%%%%%%%%%%%%%%
%
%In the following examples, we illustrate the general theory by specifying a concrete form of the Hamiltonian $H(q,P)$. Reader's familiarity with the techniques of geometric algebra and calculus on the level of Appendix \ref{sec:GAGC} is assumed.
%

%%%%%%%%%%%%%%%%%%%%%%%%%%%%%%%%%%%%%%
\section{Example: Scalar field theory} \label{sec:ExScalar}
%%%%%%%%%%%%%%%%%%%%%%%%%%%%%%%%%%%%%%

In this example, we split the configuration space $\mathcal{C}$ into a $D$-dimensional spacetime with the unit pseudoscalar $I_x$ (we will assume $D>1$), and its $N$-dimensional orthogonal complement, the space of fields, with an orthonormal basis $\{ e_a \}_{a=1}^N$, and the unit pseudoscalar $I_y$. The points in $\mathcal{C}$ then have a natural decomposition $q=x+y$.

Let us assume the following form of the Hamiltonian:
\begin{equation} \label{HamDWform}
H(q,P) = P \cdot I_x + H_{\rm DW}(q,P) ,
\end{equation}
where $H_{\rm DW}$ is the \emph{De Donder-Weyl Hamiltonian} \cite{DeDonder,Weyl,Struckmeier,Kanat1998}, which satisfies the conditions
\begin{equation} \label{DWcond}
I_x \cdot \partial_P H_{\rm DW} = 0 ~~~~~{\rm and}~~~~~
(e_b \wedge e_a) \cdot \partial_P H_{\rm DW} = 0 \quad (\forall a,b=1,\ldots,N) .
\end{equation}
Geometrically, these conditions mean that $H_{\rm DW}$ depends only on those components of the momentum $D$-vector $P$, which are composed of one vector from the $y$-space, and $D-1$ vectors from the $x$-space.

\begin{figure} 
\includegraphics[scale=1]{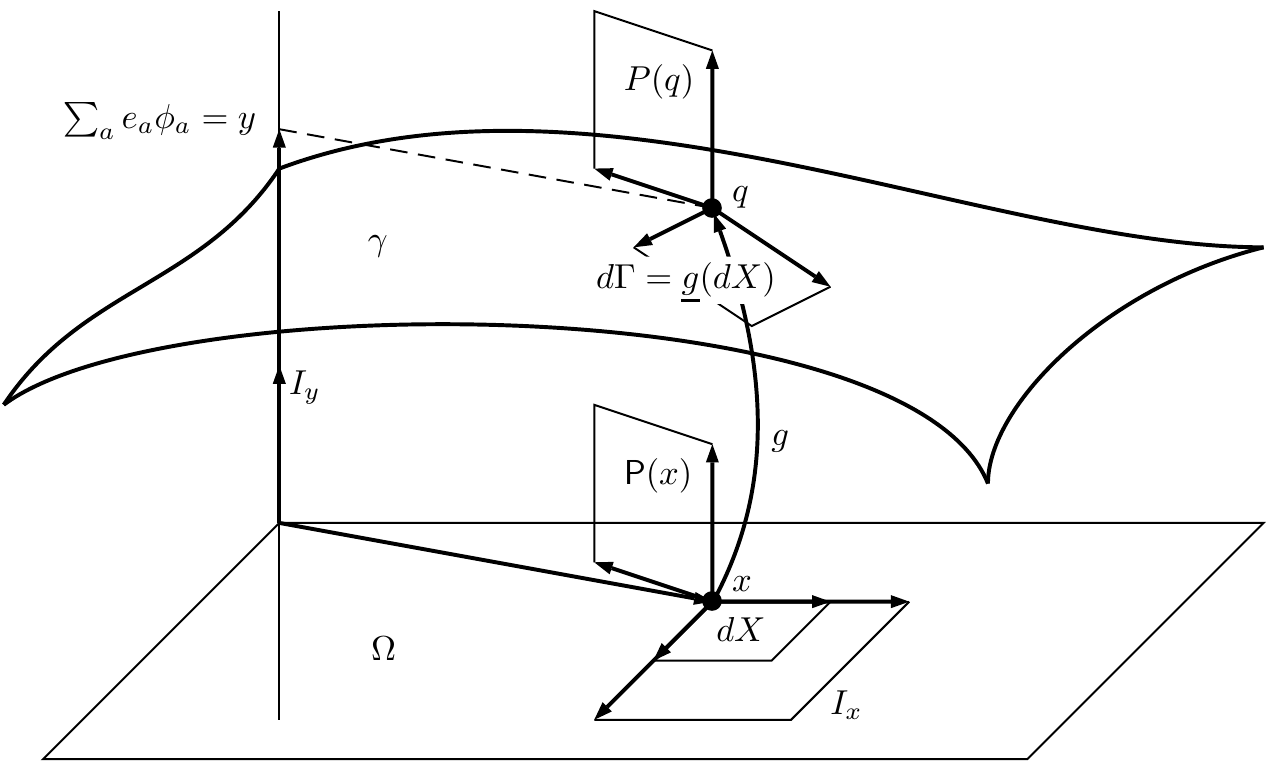}
\caption{Scalar field theory.}
\label{fig:ScField}
\end{figure}

In order to make contact with the standard theory of fields as functions defined on the spacetime, we represent the motions as 
\begin{equation} \label{MotionFunc}
\gamma=\{ x+y(x)\,|\,x\in\Omega \} ,
\end{equation} 
where $\Omega$ is a spacetime domain (see Fig.~\ref{fig:ScField}). 
The surface element of $\gamma$ is related to the oriented spacetime element $dX=|dX| I_x$ via Formula (\ref{GCgdGamma}),
\begin{equation} \label{SFdGamma}
d\Gamma = dX + (dX \cdot \partial_x) \wedge y ,
\end{equation}
where the terms with more than one $y$ have been neglected. In fact, they vanish in consequence of the second condition in (\ref{DWcond}), and the first canonical equation (\ref{CanEOM1}), which for the Hamiltonian (\ref{HamDWform}) reads
\begin{equation} \label{CanEOM1DW}
d\Gamma = \lambda I_x + \lambda \, \partial_P H_{\rm DW} .
\end{equation}
We may in addition assume that the classical momentum satisfies
\begin{equation} \label{SFMomAss}
P \cdot (e_a \wedge e_b) = 0 ~~~~~ (\forall a,b) ,
\end{equation}
as this condition has no effect on the classical motions.

%%%%%%%%%%%%%%%%%%%%%%%%%%%%%%%%%%%%%%
\subsection{De Donder-Weyl equations of motion}
%%%%%%%%%%%%%%%%%%%%%%%%%%%%%%%%%%%%%%

Comparing term by term Eqs.~(\ref{SFdGamma}) and (\ref{CanEOM1DW}), we find that
\begin{equation} \label{SFlambda}
\lambda = |dX| ,
\end{equation}
and
\begin{equation}
(I_x \cdot \partial_x) \wedge y 
= \partial_P H_{\rm DW} .
\end{equation}
The latter equation can be cast as
\begin{equation} \label{DW1}
\partial_x y 
= I_x^{-1} \partial_P H_{\rm DW} ,
\end{equation}
due to the orthogonality of the $x$- and $y$-spaces.

Formula~(\ref{GCgdGammaPartial}) can be used to ``pull" the second canonical equation~(\ref{CanEOM2}) down onto the spacetime to yield
\begin{equation}
\left[ I_x \cdot \partial_x + \big( (I_x \cdot \partial_x) \cdot \dot{\partial}_x \big) \wedge \dot{y} \right] \cdot \mathsf{P}
= (-1)^D \dot{\partial}_q H_{\rm DW}(\dot{q},\mathsf{P}) ,
\end{equation}
where we have denoted $\mathsf{P}(x) \equiv P(x+y(x))$.
``Dotting" this equation with a $y$-vector $e_a$, the second term on the left-hand side drops out due to the assumption~(\ref{SFMomAss}), and we arrive at
\begin{equation} \label{DW2}
(e_a I_x \partial_x) \cdot \mathsf{P} 
= (-1)^D e_a \cdot \partial_y H_{\rm DW} .
\end{equation}

It is now straightforward to show that, choosing an orthonormal basis of the $x$-space, the components of Eqs.~(\ref{DW1}) and (\ref{DW2}) correctly reproduce the standard equations of motion of the De Donder-Weyl Hamiltonian field theory.

%%%%%%%%%%%%%%%%%%%%%%%%%%%%%%%%%%%%%%
\subsection{Hamilton-Jacobi theory}
%%%%%%%%%%%%%%%%%%%%%%%%%%%%%%%%%%%%%%

For the Hamiltonian given by Eq.~(\ref{HamDWform}), the Hamilton-Jacobi equation~(\ref{HJeq}) reads
\begin{equation} \label{ScFieldHJ}
I_x \cdot (\partial_q \wedge S) + H_{\rm DW}(q,\partial_q \wedge S) = 0 ,
\end{equation}
where $S(q)$ is a multivector of grade $D-1$. 
This can be related to the Hamilton-Jacobi equation derived formerly by Weyl \cite{Weyl}.

To this end, let us assume that $S$ is a spacetime multivector, and define the vector $s(q)\equiv S(q) I_x$.
Taking into account the assumptions~(\ref{DWcond}), Eq.~(\ref{ScFieldHJ}) is cast as
\begin{equation}
\partial_x \cdot s + H_{\rm DW}(q,\partial_y s \, I_x^{-1}) = 0 ,
\end{equation}
which, when written out in components, is indeed the Weyl's Hamilton-Jacobi equation.

%%%%%%%%%%%%%%%%%%%%%%%%%%%%%%%%%%%%%%
\subsection{Lagrangian formulation}
%%%%%%%%%%%%%%%%%%%%%%%%%%%%%%%%%%%%%%

From now on, we shall be concerned only with a specialized form of the Hamiltonian~(\ref{HamDWform}),
\begin{equation} \label{HamScField}
H_{\rm SF}(q,P) = P \cdot I_x + \frac{1}{2} \sum_{a=1}^N \big( I_x \cdot (P \cdot e_a) \big)^2 + V(y) .
\end{equation}
Eq.~(\ref{DW1}) in this case reads
\begin{equation}
\partial_x y 
= I_x^{-1} \sum_{a=1}^N e_a \wedge (e_a \cdot \rev{\mathsf{P}}) ,
\end{equation}
where $\rev{\mathsf{P}}$ denotes the \emph{reversion} of $\mathsf{P}$ (see the definition (1.17) in \cite[Ch.~1-1]{Hestenes}). Writing the field $y$ in components, $y(x) = \sum_a e_a \phi_a(x)$, the latter equation reads
\begin{equation} \label{SFMomSubs}
\partial_x \phi_a 
= I_x^{-1} (\rev{\mathsf{P}} \cdot e_a)
= I_x (\mathsf{P} \cdot e_a) .
\end{equation}
The last equality holds in Euclidean spaces, where $I_x^{-1} = \rev{I}_x$.

At this point it is worth to note that for the Hamiltonian $H_{\rm SF}$ the extended action (\ref{ActionAugm}) can be cast, using Eqs.~(\ref{SFdGamma}) and (\ref{SFlambda}), as an integral over the spacetime domain $\Omega$,
\begin{align}
\mathcal{A}_{\rm SF}
&= \int_\Omega \left\{ \mathsf{P} \cdot \left[ dX + (dX \cdot \partial_x) \wedge y \right] - |dX| H_{\rm SF} \right\} \nonumber\\
&= \int_\Omega |dX| \left\{ (I_x \cdot \dot{\partial}_x) \cdot (\dot{y} \cdot \mathsf{P}) 
- \frac{1}{2} \sum_{a=1}^N \big( I_x \cdot (\mathsf{P} \cdot e_a) \big)^2 - V(y) \right\} .
\end{align}
Eliminating the momentum by virtue of Eq.~(\ref{SFMomSubs}), and employing the identity
\begin{equation} \label{ProjIdSc}
(a \cdot I_x) \cdot (I_x^{-1} \cdot b)
= a \cdot b ,
\end{equation}
which holds for any spacetime vectors $a$ and $b$,
we obtain
\begin{equation} \label{ActionSF}
\mathcal{A}_{\rm SF}
= \int_\Omega \mathcal{L}_{\rm SF}(\phi_a,\partial_x \phi_a) \, |dX| ,
\end{equation}
where
\begin{equation} \label{LagrSF}
\mathcal{L}_{\rm SF}(\phi_a,\partial_x \phi_a)
= \frac{1}{2} \sum_{a=1}^N (\partial_x \phi_a)^2 - V(y)
\end{equation}
is the usual Lagrangian of an $N$-component scalar field $y=(\phi_1,\ldots,\phi_N)$.
This observation justifies, a posteriori, the title of this section ``Scalar field theory".

%%%%%%%%%%%%%%%%%%%%%%%%%%%%%%%%%%%%%%
\subsection{Symmetries and the continuity equation}
%%%%%%%%%%%%%%%%%%%%%%%%%%%%%%%%%%%%%%

Equation (\ref{GCgdGammaPartial}) can be used to ``pull" the conservation law (\ref{ConsLaw}) down onto the spacetime to recover the standard form of the continuity, and relate the conserved multivectors $P \cdot v$ to the Noether currents. For this purpose, we  define $\mathsf{v}(x) \equiv v(x+y(x))$, and calculate
\begin{align}
(d\Gamma \cdot \partial_q) \cdot (P \cdot v)
&= \left[ dX \cdot \partial_x + \big( (dX \cdot \partial_x) \cdot \dot{\partial}_x \big) \wedge \dot{y} \right] \cdot (\mathsf{P} \cdot \mathsf{v}) \nonumber\\
&= (-1)^{D-1} (\partial_x \cdot dX) \cdot \left[ \mathsf{P} \cdot \mathsf{v} 
+ \dot{\partial}_x \wedge \big( \dot{y} \cdot (\mathsf{P} \cdot \mathsf{v}) \big) \right] \nonumber\\
&= |dX| (-1)^D \partial_x \cdot j(x) ,
\end{align}
where we have denoted
\begin{equation} \label{NoetherCur}
j(x) \equiv - I_x \cdot \left[ \mathsf{P} \cdot \mathsf{v} + \dot{\partial}_x \wedge \big( \dot{y} \cdot (\mathsf{P} \cdot \mathsf{v}) \big) \right] .
\end{equation}
This is the standard Noether current corresponding to the symmetry generated by the vector field $v$. In view of the conservation law~(\ref{ConsLaw}), it satisfies the continuity equation
\begin{equation}
\partial_x \cdot j(x) = 0 .
\end{equation}

\begin{figure} 
\includegraphics[scale=1]{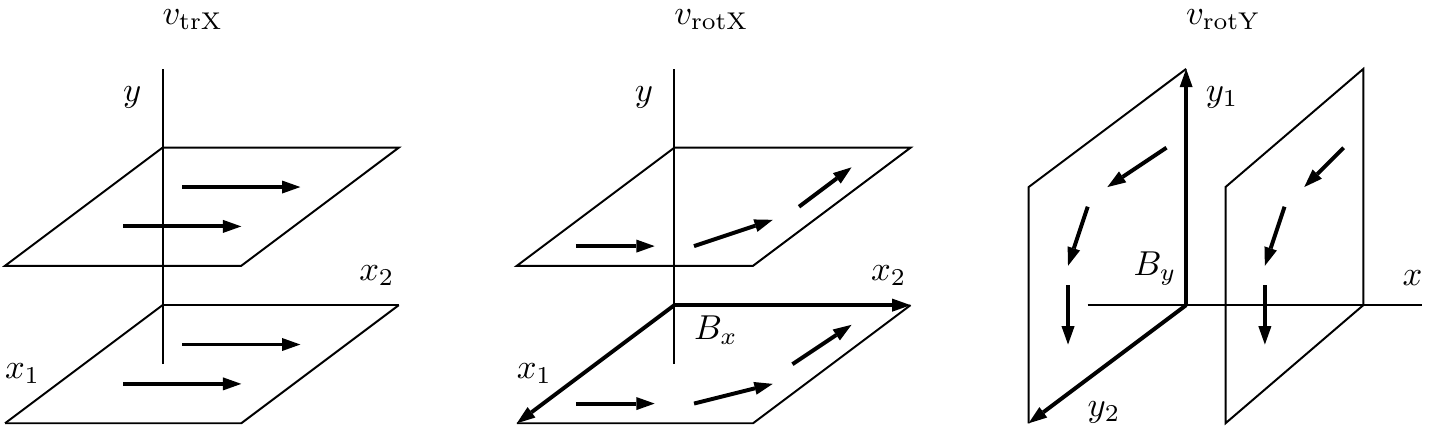}
\caption{The symmetry generators of the scalar field Hamiltonian $H_{\rm SF}$: spacetime translations $v_{\rm trX}$, spacetime rotations $v_{\rm rotX}$, and field-space rotations $v_{\rm rotY}$. The spacetime or the field space are conveniently depicted as two-dimensional planes $(x_1,x_2)$ or $(y_1,y_2)$, respectively. }
\label{fig:SFGen}
\end{figure}

We will now show that the scalar-field Hamiltonian $H_{\rm SF}$ enjoys some well known symmetries (depicted in Fig.~\ref{fig:SFGen}), and exploit the corresponding conserved currents.

\subsubsection{Translations in spacetime}

For global spacetime translations
\begin{equation}
f_{\rm trX}(q) = q + v_x ,
\end{equation}
where $v_x$ is a constant spacetime vector, the differential mapping is trivial, 
\begin{equation}
\underline{f}(a) = a ,
\end{equation}
and so is the adjoint,
\begin{equation}
\overline{f^{-1}}(P) = P .
\end{equation}
The vector $v_x$ is at the same time the generator of translations,
\begin{equation}
v_{\rm trX}(q) = v_x ,
\end{equation} 
as can be ascertained by calculating $e^{v_x \cdot \partial_q} q = f_{\rm trX}(q)$.

The transformation $f_{\rm trX}$ is, according to Eq.~(\ref{SymCrit}), a symmetry of the Hamiltonian $H_{\rm SF}$, since $H_{\rm SF}$ does not depend on $x$. The conserved quantity $P \cdot v_{\rm trX}$ is related to the Noether current $j_{\rm trX}(x)$ via Eq.~(\ref{NoetherCur}). Explicitly,
\begin{align}
j_{\rm trX} 
&= - I_x \cdot \left[ \mathsf{P} \cdot v_x + \big( (\mathsf{P} \cdot \dot{y}) \wedge  \dot{\partial}_x \big) \cdot v_x - v_x \cdot \dot{\partial}_x \, \mathsf{P} \cdot \dot{y} \right] 
\nonumber\\
&= - v_x \left[ \mathsf{P} \cdot I_x + (\mathsf{P} \cdot \dot{y}) \cdot  (\dot{\partial}_x \cdot I_x) \right]
+ v_x \cdot \dot{\partial}_x \, I_x \cdot (\mathsf{P} \cdot \dot{y}) ,
\end{align}
where we have used the fact that $I_x \wedge v_x = 0$. Substituting now for $\mathsf{P} \cdot I_x$ from the Hamiltonian constraint $H_{\rm SF} = 0$, and for $\mathsf{P} \cdot e_a$ from Eq.~(\ref{SFMomSubs}), and using the identity~(\ref{ProjIdSc}), we arrive at
\begin{align} \label{SFEnMomTensor}
j_{\rm trX}(x;v_x)
&= - v_x \left[ \frac{1}{2} \sum_{a=1}^N (\partial_x \phi_a)^2 - V(y) \right] + \sum_{a=1}^N ( v_x \cdot \partial_x \phi_a ) (\partial_x \phi_a)
\nonumber\\
&= - v_x \mathcal{L}_{\rm SF} + \sum_{a=1}^N ( v_x \cdot \partial_x \phi_a ) \frac{\partial \mathcal{L}_{\rm SF}}{\partial (\partial_x \phi_a)} .
\end{align}
This is the standard energy-momentum tensor of a scalar field with Lagrangian (\ref{LagrSF}).
In its natural geometric interpretation, $j_{\rm trX}$ is an $x$-dependent linear mapping of spacetime vectors $v_x$ to spacetime vectors $j_{\rm trX}(x;v_x)$.

\subsubsection{Rotations in spacetime}

A spacetime rotation about a point $x_0$ is defined
\begin{equation} \label{RotXdef}
f_{\rm rotX}(q) = x_0 + R_x (q - x_0) \rev{R}_x 
~~~,~~~
R_x = e^{-B_x/2} ,
\end{equation}
where $B_x$ is a constant spacetime bivector, and $R_x$ is the corresponding \emph{rotor}. The associated differential mapping is readily obtained,
\begin{equation}
\underline{f_{\rm rotX}}(a) 
= a \cdot \partial_q f_{\rm rotX}(q)
= R_x a \rev{R}_x ,
\end{equation}
and the transformation rule for the momentum is found,
\begin{equation}
\overline{f_{\rm rotX}^{-1}}(P) = R_x P \rev{R}_x .
\end{equation}
(The implementation of rotations using geometric algebra is discussed in \cite[Ch.~3-5]{Hestenes}.)

By expanding the right-hand side of the definition (\ref{RotXdef}), and comparing with the Lie series, Eq.~(\ref{LieSeries}), we find the infinitesimal generator of $f_{\rm rotX}$,
\begin{equation} \label{SFGenRotX}
v_{\rm rotX}(q) = (q - x_0) \cdot B_x = (x - x_0) \cdot B_x .
\end{equation}

In order to show that $f_{\rm rotX}$ is a symmetry of $H_{\rm SF}$, we realize that $R_x I_x \rev{R}_x = I_x$ and $R_x e_a \rev{R}_x = e_a$, and calculate
\begin{align}
H_{\rm SF}(f_{\rm rotX}(q), \overline{f_{\rm rotX}^{-1}}(P))
&= (R_x P \rev{R}_x) \cdot I_x + \frac{1}{2} \sum_{a=1}^N \big[ I_x \cdot \big((R_x P \rev{R}_x) \cdot e_a \big) \big]^2 + V(y) \nonumber\\
%%%%%%
%&= (R_x P \rev{R}_x) \cdot (R_x I_x \rev{R}_x) + \frac{1}{2} \sum_{a=1}^N \big[ (R_x I_x \rev{R}_x) \cdot \big( (R_x P \rev{R}_x) \cdot (R_x e_a \rev{R}_x) \big) \big]^2 + V(y) \nonumber\\
%%%%%%
&= P \cdot I_x + \frac{1}{2} \sum_{a=1}^N \big( I_x \cdot (P \cdot e_a) \big)^2 + V(y) = H_{\rm SF}(q,P) .
\end{align}

Since $v_{\rm rotX}$ is a spacetime vector, it is easy to find an explicit relation between $\mathsf{P} \cdot v_{\rm rotX}$ and the corresponding Noether current $j_{\rm rotX}$. We simply replace in Eq.~(\ref{SFEnMomTensor}) $v_x$ by $\mathsf{v}_{\rm rotX}$:
\begin{equation} \label{SFAngMomTensor}
j_{\rm rotX}(x;B_x,x_0) 
= j_{\rm tr}\big(x; \mathsf{v}_{\rm rotX} \big)
= j_{\rm tr}\big(x;(x-x_0)\cdot B_x \big) .
\end{equation}
This is the angular momentum tensor corresponding to the energy-momentum tensor $j_{\rm tr}$. Geometrically, $j_{\rm rotX}$ is an $x$-dependent linear mapping, with a parameter $x_0$, that maps spacetime bivectors $B_x$ to spacetime vectors $j_{\rm rotX}(x;B_x,x_0)$ (c.f. Ch.~13.1 in Ref.~\cite{DoranLas}).

\subsubsection{Rotations in field space}

Finally, let us consider rotations of the form
\begin{equation}
f_{\rm rotY}(q) = R_y q \rev{R}_y 
~~~,~~~
R_y = e^{-B_y/2} ,
\end{equation}
where $B_y$ is a constant bivector from the field space, i.e., $B_y \cdot I_y = B_y I_y$. The corresponding differential reads
\begin{equation}
\underline{f_{\rm rotY}}(a) 
= a \cdot \partial_q f_{\rm rotY}(q)
= R_y a \rev{R}_y ,
\end{equation}
and the momentum transforms as
\begin{equation}
\overline{f_{\rm rotY}^{-1}}(P) = R_y P \rev{R}_y .
\end{equation}
The generator of the field-space rotations is found in the same way as the generator of the spacetime rotations (\ref{SFGenRotX}), 
\begin{equation}
v_{\rm rotY}(q) = q \cdot B_y = y \cdot B_y .
\end{equation}

The Hamiltonian $H_{\rm SF}$ transforms under $f_{\rm rotY}$ as follows (note that $R_y I_x \rev{R}_y = I_x$):
\begin{equation} \label{SFHamRotY}
H_{\rm SF}(f_{\rm rotY}(q), \overline{f_{\rm rotY}^{-1}}(P))
= P \cdot I_x + \frac{1}{2} \sum_{a=1}^N \big[ I_x \cdot \big(P \cdot (\rev{R}_y e_a R_y) \big) \big]^2 + V(R_y y \rev{R}_y) .
\end{equation}
If we assume that $V(R_y y \rev{R}_y) = V(y)$, which is fulfilled, for example, when the potential $V$ depends only on $y^2 = \sum_a \phi_a^2$, then the right-hand side of Eq.~(\ref{SFHamRotY}) is equal to $H_{\rm SF}(q,P)$, and hence $f_{\rm rotY}$ is a symmetry of $H_{\rm SF}$.  Note that the second term in $H_{\rm SF}$ is invariant under a change of the orthonormal basis of the $y$-space, $e_a \rightarrow e_a'=\rev{R}_y e_a R_y$, as can be easily ascertained.

The vector field $v_{\rm rotY}$ lies entirely in the $y$-space. Therefore, owing to the assumption~(\ref{SFMomAss}), the second term in expression (\ref{NoetherCur}) for the Noether current drops out, and we obtain
\begin{equation}
j_{\rm rotY}
= -I_x \cdot (\mathsf{P} \cdot \mathsf{v}_{\rm rotY})
= - \sum_{a=1}^N I_x \cdot (\mathsf{P} \cdot e_a) \, (e_a \wedge y) \cdot B_y
\end{equation}
A substitution for the momentum from Eq.~(\ref{SFMomSubs}) then yields
\begin{equation}
j_{\rm rotY}(x;B_y)
= \dot{\partial}_x \, (y \wedge \dot{y}) \cdot B_y
= \sum_{a,b=1}^N (e_a \wedge e_b) \cdot B_y \, \phi_a \partial_x \phi_b .
\end{equation}
The Noether current $j_{\rm rotY}$ is an $x$-dependent linear mapping of field-space bivectors $B_y$ to spacetime vectors $j_{\rm rotY}(x;B_y)$.

%%%%%%%%%%%%%%%%%%%%%%%%%%%%%%%%%%%%%%
\section{Example: String theory} \label{sec:ExString}
%%%%%%%%%%%%%%%%%%%%%%%%%%%%%%%%%%%%%%

Probably the simplest nontrivial Hamiltonian, which preserves the full symmetry of the configuration space $\mathcal{C}$, is 
\begin{equation}
H_{\rm Str} = \frac{1}{2}(|P|^2 - \Lambda^2) ,
\end{equation}
where $\Lambda >0$ is a scalar constant, and $|P|=\sqrt{\rev{P}\cdot P}$ is the magnitude of $P$.
This Hamiltonian described the dynamics of a relativistic particle (for $D=1$), a string (for $D=2$), or a higher-dimensional membrane (for $D>2$) that propagates in a Euclidean spacetime $\mathcal{C}$. The corresponding worldlines (or worldsheets) are identified with the motions $\gamma$.

%%%%%%%%%%%%%%%%%%%%%%%%%%%%%%%%%%%%%%
\subsection{Equations of motion}
%%%%%%%%%%%%%%%%%%%%%%%%%%%%%%%%%%%%%%

The first canonical equation (\ref{CanEOM1}) takes the form 
\begin{equation} \label{StringEOM1}
d\Gamma = \lambda \rev{P} ,
\end{equation}
which, when substituted into the Hamiltonian constraint (\ref{CanEOM3}), fixes the absolute value of the Lagrange multiplier $\lambda$,
\begin{equation} \label{StringEOM3}
|d\Gamma| = |\lambda| \Lambda .
\end{equation}
Furthermore, substituting Eq.~(\ref{StringEOM1}) into the second canonical equation of motion (\ref{CanEOM2}), dividing by $\lambda$, and using Eq.~(\ref{StringEOM3}), we find
\begin{align} \label{StringEOM}
I_\gamma \cdot \partial_q \, I_\gamma &= 0 ~~~~~ ({\rm for}~D=1) ,
\nonumber \\
(I_\gamma \cdot \partial_q ) \cdot I_\gamma &= 0 ~~~~~ ({\rm for}~D>1),
\end{align}
where $I_\gamma \equiv d\Gamma / |d\Gamma|$ is the unit pseudoscalar of the surface $\gamma$. This equation has a simple geometric interpretation. It entails vanishing of the \emph{mean curvature} of the surface $\gamma$, or, more generally, of its  \emph{spur} vector (see Ref.~\cite[Ch. 4-4]{Hestenes}).

%%%%%%%%%%%%%%%%%%%%%%%%%%%%%%%%%%%%%%
\subsection{Nambu-Goto action}
%%%%%%%%%%%%%%%%%%%%%%%%%%%%%%%%%%%%%%

Eqs.~(\ref{StringEOM1}) and (\ref{StringEOM3}) allow us to eliminate $P$ and $\lambda$, and rewrite the action (\ref{Action}) in terms of $d\Gamma$ only,
\begin{equation} \label{NambuGotoAction}
\mathcal{A}_{\rm Str}
= \int_\gamma P \cdot d\Gamma 
= \int_\gamma \frac{1}{\lambda} |d\Gamma|^2
= \pm\,\Lambda \int_\gamma |d\Gamma| ,
\end{equation}
where ``$\pm$" is the sign of $\lambda$.
This is the Euclidean Nambu-Goto action of the bosonic string theory \cite{Zwiebach}. It is proportional to the volume of the worldsheet $\gamma$, with $\Lambda$ playing the role of the string tension.
% (taking a unit speed of light). 

The extremals of the action $\mathcal{A}_{\rm Str}$, i.e., the solutions of Eq.~(\ref{StringEOM}), minimize their volume for a given fixed boundary. Therefore, they are called \emph{minimal surfaces} in the mathematical literature \cite{Osserman}. 

If assume that the worldsheets are nearly flat, and represent them in the same way as the scalar field, Eq.~(\ref{MotionFunc}), then the string action is cast as
\begin{equation}
\mathcal{A}_{\rm Str}
\approx \pm\,\Lambda \int_\Omega |dX| | I_x + (I_x \cdot \partial_x) \wedge y | 
%= \pm\,\Lambda \int_\Omega |dX| \sqrt{1 + \sum_{a=1}^N (\partial_x \phi_a)^2} 
= \pm\,\Lambda \int_\Omega |dX| \left[ 1 + \frac{1}{2}\sum_{a=1}^N (\partial_x \phi_a)^2 \right]  ,
\end{equation}
where $\phi_a \equiv e_a \cdot y$, and the terms of order greater than $(\partial_x \phi_a)^2$ have been neglected.
A comparison with the scalar-field action $\mathcal{A}_{\rm SF}$, Eq.~(\ref{ActionSF}), yields
\begin{equation}
\mathcal{A}_{\rm Str} 
\approx \pm\,\Lambda\, \mathcal{A}_{\rm SF}|_{V=0} \pm\,\Lambda \int_\Omega |dX| ,
\end{equation}
and hence we conclude that the string theory for slowly varying worldsheets essentially reduces to a potential-free massless scalar field theory.

%%%%%%%%%%%%%%%%%%%%%%%%%%%%%%%%%%%%%%
\subsection{Hamilton-Jacobi theory}
%%%%%%%%%%%%%%%%%%%%%%%%%%%%%%%%%%%%%%

In this example, the Hamilton-Jacobi equation (\ref{HJeq}) takes a particularly compact form
\begin{equation} \label{StringHJ}
| \partial_q \wedge S | = \Lambda ,
\end{equation}
reproducing the result of Ch.~7 in Ref. \cite{Kastrup}.

%%%%%%%%%%%%%%%%%%%%%%%%%%%%%%%%%%%%%%
\subsection{Physical motions of a relativistic particle}
%%%%%%%%%%%%%%%%%%%%%%%%%%%%%%%%%%%%%%

For the moment, let us focus on the case $D=1$, which describes a relativistic particle in the Euclidean spacetime. We will present two methods for finding the physical motions $\gamma_{\rm cl}$. 

First, suppose that two points, $q_0$ and $q$, lie on $\gamma_{\rm cl}$, multiply the equation of motion~(\ref{StringEOM}) by $|d\Gamma|$, and integrate along $\gamma_{\rm cl}$ from $q_0$ to $q$. The fundamental theorem of calculus implies that
\begin{equation}
I_\gamma(q) - I_\gamma(q_0) = 0 ,
\end{equation}
i.e., $I_\gamma$ is constant along $\gamma_{\rm cl}$. The physical motions are therefore straight lines in $\mathcal{C}$, 
\begin{equation}
\gamma_{\rm cl} = \{ q = v \tau + q_0 \,|\, \tau \in \mathbb{R} \} ,
\end{equation}
where $q_0 \in \mathcal{C}$ and $v$ is an arbitrary constant vector.

The second method makes use of a family of solutions of the Hamilton-Jacobi equation (\ref{StringHJ}). Take, for example,
\begin{equation}
S(q;q_0) = \Lambda |q-q_0| .
\end{equation}
The derivative of $S$ with respect to $q_0$ yields, according to Formula (\ref{HJconserved}), the conserved quantities
\begin{equation}
\partial_{q_0} S = - \Lambda \frac{q-q_0}{|q-q_0|} .
\end{equation}
The physical motion are then given by
\begin{equation}
\gamma_{\rm cl} = \left\{
q \, \bigg| \, \frac{q-q_0}{|q-q_0|} = v
\right\} ,
\end{equation}
where $v$ is an arbitrary constant unit vector.

%%%%%%%%%%%%%%%%%%%%%%%%%%%%%%%%%%%%%%
\subsection{Symmetries and conserved quantities}
%%%%%%%%%%%%%%%%%%%%%%%%%%%%%%%%%%%%%%

For the Hamiltonian $H_{\rm Str}$, the infinitesimal symmetry condition, Eq.~(\ref{SymCritInfsm}), reads
\begin{equation} \label{StrSym}
\big( \dot{\partial}_q \wedge (\dot{v} \cdot P) \big) \cdot \rev{P}
= 0 .
\end{equation}
This has to be satisfied for all constant $D$-vectors $P$. Observe that the left-hand side is equal to
\begin{equation}
\frac{1}{2} \left[ \dot{\partial}_q \wedge (\dot{v} \cdot P) + \dot{v} \wedge (\dot{\partial}_q \cdot P) \right] \cdot \rev{P}
= \frac{1}{2} \sum_{j=1}^{N+D} ( \partial_q v \cdot e_j + e_j \cdot \partial_q v ) \cdot \big( (e_j \cdot P) \cdot \rev{P} \big) ,
\end{equation}
where the $e_j$'s form an orthonormal basis of the configuration space $\mathcal{C}$.
%Differentiation with respect to $P$ yields
%\begin{equation}
%\dot{\partial}_q \wedge (\dot{v} \cdot P)
%+ (P \cdot \partial_q) \wedge v = 0 ,
%\end{equation}
The solution of Eq.~(\ref{StrSym}) is therefore a vector field $v$, for which 
\begin{equation} \label{SymGenStr}
a \cdot \partial_q v = - \partial_q v \cdot a 
\end{equation}
holds for all constant vectors $a$. 
%In view of definitions (\ref{GAdifMap}) and (\ref{GCadjoint}), we can write succinctly $\uline{v}=-\oline{v}$, i.e., $\uline{v}$ is a skew-symmetric linear mapping.
Taking the curl, the right-hand side vanishes, and we find 
\begin{equation}
a\cdot\partial_q \, \partial_q \wedge v = 0 ,
\end{equation}
which implies
\begin{equation}
\partial_q \wedge v = 2 B_0 ,
\end{equation}
where $B_0$ is a constant bivector. Moreover, note that from Eq.~(\ref{SymGenStr}) follows that 
\begin{equation}
a \cdot ( \partial_q \wedge v ) = 2 a \cdot \partial_q v ,
\end{equation}
and hence
\begin{equation}
a \cdot \partial_q v - a \cdot B_0 = a \cdot \partial_q (v - q \cdot B_0) = 0 ,
\end{equation}
from which we finally obtain an expression for the symmetry generator $v$,
\begin{equation}
v(q) = q \cdot B_0 + v_0 ,
\end{equation}
where $v_0$ is a constant vector.

The vector field $v$ is composed of two terms, the translation generator 
\begin{equation}
v_{\rm tr}(q) = v_0 ,
\end{equation}
and the rotation generator
\begin{equation}
v_{\rm rot}(q) = q \cdot B_0 .
\end{equation}
The corresponding finite symmetry transformations can be obtained directly from the Lie series, Eq.~(\ref{LieSeries}), and read (setting $\tau = 1$)
\begin{equation}
f_{\rm tr}(q) = q + v_0 ,
\end{equation}
and
\begin{equation}
f_{\rm rot}(q) 
= q + q \cdot B_0 + \frac{1}{2!} \big( q \cdot B_0 \big) \cdot B_0 + \ldots
= e^{- B_0/2} \, q \, e^{B_0/2} , 
\end{equation}
respectively.

In view of Eqs.~(\ref{StringEOM1}) and (\ref{StringEOM3}), the conserved quantities take the form
\begin{equation}
P \cdot v = \pm \Lambda \, \rev{I}_\gamma \cdot v ,
\end{equation}
where ``$\pm$" is the sign of $\lambda$.

%%%%%%%%%%%%%%%%%%%%%%%%%%%%%%%%%%%%%%
\section{Conclusion and outlook} \label{sec:Conclusion}
%%%%%%%%%%%%%%%%%%%%%%%%%%%%%%%%%%%%%%

In this article we studied and developed the formulation of classical field theories proposed in Ref.~\cite[Ch.~3]{RovelliQG}. This formulation is based on the Hamiltonian constraint $H(q,P)=0$ between the partial observables $q$, and the generalized momentum $P$, and the fields are viewed as unparametrized submanifolds of the configuration space.

Starting from the variational principle of Sec.~\ref{sec:VarPrinc}, we derived the canonical equations of motion (\ref{CanEOM}), and, subsequently, deduced the local Hamilton-Jacobi equation (\ref{HJeq}). These results generalize to the field theory the respective notions from the Hamiltonian particle mechanics. In Sec.~\ref{sec:Sym}, we discussed transformations of the configuration space, and identified the symmetry transformations by the condition (\ref{SymCrit}). In the ensuing section, taking into account the canonical equations of motion, symmetries were shown to imply conservation laws, Eq.~(\ref{ConsLaw}), thus establishing a Hamiltonian field-theoretical version of the Noether theorem. The simple form of the conserved quantities $P \cdot v$, where $v$ is the vector field that generates the symmetry, clarifies the physical significance of the momentum multivector $P$.

With two ensuing examples, we showed that scalar fields and strings can both be accommodated within our formalism. One only has to take the appropriate Hamiltonian constraint. In fact, as we also demonstrated, the scalar field theory is a limiting case of the string theory, in a similar way in which the non-relativistic particle mechanics is a limiting case of the relativistic mechanics. To make contact with the standard treatment, we showed that our Hamiltonian constraint approach to the scalar field theory leads to the De Donder-Weyl formalism. Moreover, we expressed the energy-momentum tensor of the scalar field in terms of the conserved multivector $P \cdot v$ to argue that the latter is a more primitive, and therefore more fundamental, object.

The Hamiltonian formalism is especially important when it comes to quantization. In mechanics, the momentum is promoted to a differential operator, and the Hamilton-Jacobi equation is replaced by the Schr\"{o}dinger equation. It is desirable to have an analogous quantization scheme also for the field theory, which is currently most commonly quantized using Lagrangians and Feynman path integrals. Within the De Donder-Weyl field theory, quantum momentum operators and a Schr\"{o}dinger-like equation have already been proposed \cite{Kanat1999,Kanat2013}. We would like to make use of these lessons to develop an analogous formulation of the quantum field theory, based on the more general Hamiltonian constraint approach.

%%%%%%%%%%%%%%%%%%%%%%%%%%%%%%%%%%%%%%
\subsection*{Acknowledgement}
%%%%%%%%%%%%%%%%%%%%%%%%%%%%%%%%%%%%%%
The author would like to thank Igor Kanatchikov for valuable discussions, and the following institutions for financial support: Czech Technical University in Prague, Grant SGS13/217/OHK4/3T/14, Czech Science Foundation (GA\v{C}R), Grant GA14-07983S, and Deutsche Forschungsgemeinschaft (DFG), Grant KL 256/54-1.

\appendix

%%%%%%%%%%%%%%%%%%%%%%%%%%%%%%%%%%%%%%
\section{Transformations and induced mappings}
%%%%%%%%%%%%%%%%%%%%%%%%%%%%%%%%%%%%%%
\label{sec:GAGC}

Let $f:\mathcal{C}\rightarrow\mathcal{C}$ be a diffeomorphism relating points in the configuration space, and consider the directional derivative
\begin{equation}
\uline{f}(a;q)
\equiv a \cdot \partial_q f(q)
= \lim_{\eps\rightarrow 0} \frac{f(q+\eps a)-f(q)}{\eps} .
\end{equation}
$\uline{f}$ is the \emph{differential}, a $q$-dependent linear mapping of vectors at a point $q$ to vectors at $f(q)$. It can be extended to an \emph{outermorphism} acting on the whole geometric algebra by demanding the property
\begin{equation}
\uline{f}(A \wedge B) = \uline{f}(A) \wedge \uline{f}(B)
\end{equation}
for all multivectors $A$ and $B$.
The \emph{adjoint} of $\uline{f}$, denoted $\oline{f}$, is defined via the relation
\begin{equation}
\uline{f}(a) \cdot b
= a \cdot \oline{f}(b) .
\end{equation}

Let us specialize to infinitesimal diffeomorphisms
\begin{equation} \label{GCinsfmDiffeo}
f(q) = q + \delta q(q) \quad,\quad \delta q(q) \equiv \eps v(q) ,
\end{equation}
where $v$ is a vector field on the configuration space $\mathcal{C}$.
The action of the differential on an $r$-blade $A_r = \blade{a}{r}$ is given by
\begin{equation}
\uline{f}(A_r) 
= (a_1 + \eps \, a_1 \cdot \partial_q v) \wedge \ldots \wedge (a_r + \eps \, a_r \cdot \partial_q v)
= A_r + \eps (A_r \cdot \partial_q) \wedge v + O(\eps^2) .
\end{equation}
For the adjoint outermorphism, we find
\begin{equation}
\oline{f}(B_r) \cdot A_r
= B_r \cdot \uline{f}(A_r) 
= B_r \cdot A_r + \eps \big( \dot{\partial}_q \wedge (\dot{v} \cdot B_r) \big) \cdot A_r + O(\eps^2) .
\end{equation} 
By the linearity of the above expressions, we therefore conclude that for an arbitrary multivector $A$,
\begin{align} \label{GCinfsmTr}
\uline{f}(A) &\approx A + \eps (A \cdot \partial_q) \wedge v , \nonumber\\
\oline{f}(A) &\approx A + \eps \, \dot{\partial}_q \wedge (\dot{v} \cdot A) ,
\end{align}
up to the first order in $\eps$. In this approximation, the inverse of $f$ reads $f^{-1}(q) = q - \eps v(q)$, and so we immediately obtain also
\begin{align} \label{GCinfsmInvTr}
\uline{f}^{-1}(A) &\approx A - \eps (A \cdot \partial_q) \wedge v , \nonumber\\
\oline{f^{-1}}(A) &\approx A - \eps \, \dot{\partial}_q \wedge (\dot{v} \cdot A) .
\end{align}

Now, let us briefly consider an example of a mapping between two different manifolds. In the scalar field theory, Sec.~\ref{sec:ExScalar}, we represented the motions $\gamma$ by the functions
\begin{equation}
g(x) = x + y(x) .
\end{equation}
A spacetime blade $A_r=\blade{a}{r}$ is then mapped by the associated outermorphism 
\begin{equation}
\underline{g}(a;x) = a \cdot \partial_x g(x) = a + a \cdot \partial_x y(x)
\end{equation}
to a blade 
\begin{equation} \label{GCgdifmap}
\underline{g}(A_r) 
= (a_1+a_1\cdot\partial_x y) \wedge \ldots \wedge (a_r+a_r\cdot\partial_x y) 
= A_r + (A_r \cdot \partial_x ) \wedge y + \ldots
\end{equation}
in the tangent algebra of $\gamma$. Here, $\partial_x$ denotes the vector derivative with respect to a spacetime point, and the ellipsis gathers the terms with two and more $y$'s. 

Formula~(\ref{GCgdifmap}) can be applied to express the oriented surface element of $\gamma$ as
\begin{equation} \label{GCgdGamma}
d\Gamma = \uline{g}(dX) 
= dX + (dX \cdot \partial_x ) \wedge y + \ldots .
\end{equation}
Moreover, from the chain rule for differentiation
\begin{equation}
a \cdot \partial_x F(g(x))
= \uline{g}(a) \cdot \partial_q F(q)
= a \cdot \oline{g}(\partial_q) F(q) , 
\end{equation}
we find that
\begin{equation} \label{GCgdGammaPartial}
d\Gamma \cdot \partial_q 
= \uline{g}(dX) \cdot \oline{g^{-1}}(\partial_x)
= \uline{g}(dX \cdot \partial_x)
= dX \cdot \partial_x + \big( (dX \cdot \partial_x) \cdot \dot{\partial}_x \big) \wedge \dot{y} + \ldots ,
\end{equation}
where $dX=|dX|I_x$, and we have used the identities~(1.14) from Ref.~\cite[Ch.~3]{Hestenes}. 

The differential operator $dX \cdot \partial_x$ acts on all functions to its right. Whether these include also $y(x)$ in the second, or higher, term on the right-hand side of Eq.~(\ref{GCgdGammaPartial}) has no effect, since $(dX \cdot \partial_x) \cdot \dot{\partial}_x = dX \cdot (\partial_x \wedge \dot{\partial}_x)$, and $\partial_x \wedge \partial_x = 0$.

%%%%%%%%%%%%%%%%%%%%%%%%%%%%%%%%%%%%%%

\end{document}